\documentclass[apj]{emulateapj}
\slugcomment{{\sc Accepted for publication in the ApJ}}

\newcommand{\sgr}{SGR~0501+4516~}
\newcommand{\sgrnos}{SGR~0501+4516}
\newcommand{\degree}{$^{\rm o}$}

\begin{document} 
\vspace{0.8 in} 

\title{Spatial, Temporal and Spectral Properties of X-ray Emission from the Magnetar \sgrnos} 

\author{Ersin {G\"o\u{g}\"u\c{s}}\altaffilmark{1}, 
Peter~M.~Woods\altaffilmark{2},
Chryssa~Kouveliotou\altaffilmark{3}
Yuki~Kaneko\altaffilmark{1},
Bryan~M.~Gaensler\altaffilmark{4},
Shami~Chatterjee\altaffilmark{5}
}
 
\altaffiltext{1}{Sabanc\i~University, Orhanl\i$-$Tuzla, \.{I}stanbul 34956 Turkey}
\altaffiltext{2}{Corvid Technologies, 689 Discovery Drive, Huntsville, AL 35806, USA}
\altaffiltext{3}{Space Science Office, VP-62, NASA/Marshall Space Flight Center, Huntsville, AL 35812, USA} 
\altaffiltext{4}{Sydney Institute for Astronomy, School of Physics A29, The University of Sydney, NSW 2006, Australia} 
\altaffiltext{5}{Department of Astronomy, Cornell University, Ithaca, NY 14853, USA} 


\begin{abstract} 

\sgr was discovered with the {\it Swift} satellite on 2008 August 22, after it emitted a series of very energetic bursts. Since then, the source was extensively monitored with {\it Swift}, the Rossi X-ray Timing Explorer ({\it RXTE}) and observed with {\it Chandra} and {\it XMM-Newton}, providing a wealth of information about its outburst behavior and burst induced changes of its persistent X-ray emission. Here we report the most accurate location of \sgr (with an accuracy of 0.11$\arcsec$) derived with {\it Chandra}. Using the combined {\it RXTE}), {\it Swift}/X-ray Telescope, {\it Chandra} and {\it XMM-Newton} observations we construct a phase connected timing solution with the longest time baseline ($\sim$240 days) to date for the source. We find that the pulse profile of the source is energy dependent and exhibits remarkable variations associated with the \sgr bursting activity. We also find significant spectral evolution (hardening) of the source persistent emission associated with bursts. Finally, we discuss the consequences of the \sgr proximity to the supernova remnant, SNR G160.9+2.6 (HB9).

\end{abstract} 


\section{Introduction} 

Soft Gamma Repeaters (SGRs) are intriguing manifestations of neutron stars. They are most often discovered in hard X-rays/soft $\gamma$-rays, when they (repeatedly) emit clusters of energetic bursts with peculiar properties; these episodes are characterized by a plethora of `short bursts' that usually last only a fraction of a second but emit extremely large energies of 10$^{37}$$-$10$^{40}$ erg. On rare occasions, SGRs emit relatively more energetic, so called `intermediate events', which are 1-2 orders of magnitude larger in energy, last longer ($\sim$1-4 s) and often exhibit an extended tail. On extremely rare instances (only three have been detected thus far), SGRs produce `giant flares', which are even more energetic (10$^{44}$$-$10$^{47}$ erg) and have distinctive and similar morphologies: a very bright and spectrally hard initial short spike ($<0.5$-s) that is followed by a longer (300$-$600 s) tail clearly modulated at the spin frequency of the originating source \citep[see][for a review]{woods06}.

Along with their bursting behavior, SGRs and their close relatives, Anomalous X-ray Pulsars (AXPs) display interesting persistent X-ray emission properties: they pulse with spin periods in the narrow range of 2$-$12 s. They all slow down at rather large spindown rates ($\sim$10$^{-11}$ s/s), and their inferred dipole magnetic fields are extremely strong (B$_{\rm d}$\footnotemark \footnotetext{B$_{\rm d}$ $\propto$ P$\dot{\rm P}$, assuming that the neutron star slows down via magnetic dipole radiation.} $\sim$ 10$^{14}$$-$10$^{15}$ G; \citealt{kouv98}). Moreover, their X-ray intensities vary, usually in the form of a rapid increase associated with the onset of a major bursting episode, followed by a gradual decrease lasting months to years (see e.g., \citealt{woods01, espo08, woods04}).

SGRs and AXPs are now classified as magnetars -- neutron stars that are powered by their extremely strong magnetic fields (B $\gtrsim$10$^{14}$ G; \citealt{dt92}). The magnetar model attributes the origin of the energetic SGR events to cracking of the solid neutron star crust, when strained by the drifting strong magnetic field: local fractures leading to short bursts \citep{td95} and cracking at global scales giving rise to giant flares \citep{td95, td01}. The persistent emission of magnetars likely originates from the stellar surface heated by the decay of the strong magnetic field \citep{td96, ozel01, hl01} and from resonant Compton scattering of soft photons by magnetospheric currents driven by twists in the evolving magnetic field \citep{tlk02}. 

\sgr was discovered on 2008 August 22, when four short and soft bursts from the source triggered the Burst Alert Telescope ({\it BAT}) onboard NASA's {\it Swift} satellite \citep{bart08}. To confirm its magnetar nature, we initiated our Target of Opportunity (ToO) observations with the Rossi X-ray Timing Explorer ({\it RXTE}). The first {\it RXTE}/Proportional Counter Array ({\it PCA}) observation took place on August 22 for a net exposure of 484 s. Timing analysis of this short stretch of data revealed coherent pulsations with a period of 5.76 s \citep{gogus08}. Subsequent observations of the new source with {\it RXTE} and the {\it Swift}/X-ray Telescope \citep[{\it XRT};][]{burr05} allowed us to determine the spindown rate of \sgrnos, and therefore, to firmly establish it as a magnetar \citep{woods08, rea09}. 

Following the onset of its outburst episode, the source flux decayed exponentially with an {\it  e}-folding time of 23.8 days \citep{rea09}. Using a subset of the {\it Swift}/XRT observations we also employ here and in addition their {\it XMM-Newton} and {\it Suzaku} observations, \citet{rea09} determined a pulse ephemeris, which showed evidence of strong negative $\ddot{\rm P}$. The latter was interpreted as the recovery to a secular spindown, which might have increased in conjunction with the outburst onset.  \citet{rea09} also analysed an archival ROSAT observation of the source and found that the X-ray flux of \sgr in 1992 was about 80 times lower. Recently, using {\it Suzaku} observations of \sgr obtained four days after the outburst onset, \citet{enoto10} reported that the persistent emission of the source extends up to about 70 keV. Finally, besides {\it Swift}/BAT, \sgr also triggered the Gamma ray Burst Monitor (GBM) onboard the Fermi Gamma-ray Space Telescope ({\it FGST}) 26 times. Both instruments recorded burst activity from the source until about 2008 September 3. Detailed analysis of the GBM data is currently underway (L. Lin et al., in preparation).
 
We present here spatial, and long term temporal and spectral characteristics of the persistent X-ray emission of \sgr using wideband X-ray observations during three states of the source: (i) its burst active episode, (ii) its transition to quiescence and finally, (iii) its quiescent phase. In Section 2, we describe the observations and data used in our study. In Section 3, we present the results of our detailed analyses. We discuss the interpretation of our results in Section 4.

\section{Observations}

\subsection{{\it RXTE}}

We monitored the source with {\it RXTE} until 2008 August 30 in densely spaced pointings (consecutive intervals ranging between $0.1 - 0.9$ days) followed by longer intervals as the source became fainter (spaced 2$-$7 days apart). Exposure times of these observations varied between $\sim$1 ks (in two cases) and $\sim$11.6 ks (in one case), while the great majority lasted $\sim$2.5 ks. Overall, we acquired a total exposure of 82.5 ks in 29 {\it RXTE} pointings from 2008 August 22 (MJD 54700) to October 14 (MJD 54753). We used data in collected with the {\it RXTE/PCA} only. The number of operating proportional counter units (PCUs) varied between 1 and 3, with a median value of 2 during our pointings. 
For each observation, we applied our burst search algorithm (see \citet{gogus00} for its methodology) to the lightcurve with 0.125 s time bins to identify short events. We then filtered out the times of short bursts identified to obtain a burst-free event list for timing studies. We finally converted all event arrival times to the Solar system barycenter. 

\subsection{{\it Swift}} 

\begin{figure}
\vspace{0.0in}
\centerline{
\epsscale{1.1}
\plotone{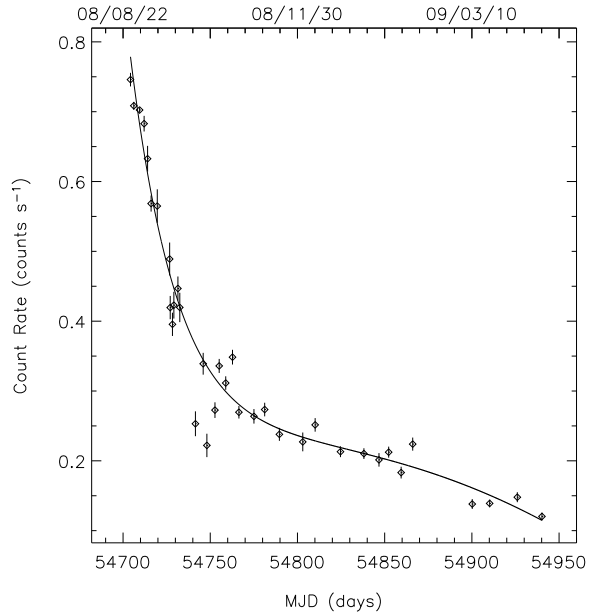}}
\vspace{0.1in}
\caption{{\it Swift}/{\it XRT} X-ray (0.3$-$10 keV) light curve of \sgrnos. The solid line is the best fit exponential function. The dates shown on top are in YY/MM/DD format. 
\label{fig:xrt_lc}}
\end{figure}

\sgr was monitored with the {\it Swift}/{\it XRT} starting 2008 August 26 (MJD 54704) until 2009 April 19 (MJD 54940) for a total exposure of 436 ks, providing the longest baseline coverage of the source to date. In Figure \ref{fig:xrt_lc} we present the long term X-ray light curve variations of \sgr as seen in the 0.3$-$10 keV range with {\it Swift}/{\it XRT}. The solid line is obtained by fitting an exponential function, whose time reference is the time of the first burst observed on August 22 (i.e., 54700.529 MJD), that yields an $e$-folding time of 27.9$\pm$2.5 days\footnotemark \footnotetext{Note that uncertainties reported throughout the paper are 1$\sigma$ unless otherwise indicated.}. This value is notably consistent (within 1.6$\sigma$) with that reported by \citet{rea09} using the first 160 days of data coverage of the source.  

We used all {\it XRT} observations that were performed in Windowed Timing (WT) mode. For each pointing we extracted source events from a 15\arcsec segment of the one dimensional WT image, centered at the pixel with the highest rate. The background events were selected from source free intervals on both sides of the source extraction interval. We then generated and filtered the light curves of each {\it XRT} pointing to ensure that there are no short SGR bursts in the persistent emission data included in our timing analysis. Finally, we transformed the arrival time of each event to the Solar system barycenter for timing studies. 

We also employed additional {\it Swift}/{\it XRT} observations in Photon Counting (PC) mode that took place around 2009 October 9, December 6 and 2010 February 20 for 13.2 ks, 13.4 ks and 11.6 ks, respectively. We performed spectral analysis with these data sets and derived the source flux at late stages of the outburst decay.

\subsection{{\it Chandra X-ray Observatory}} 

We initiated our target of opportunity observations with the {\it Chandra}/Advanced CCD Imaging Spectrometer \citep[{\it ACIS};][]{garm03} in Continuous Clocking (CC) mode on 2008 August 26 (MJD 54704) for an effective exposure of 36.5 ks. The source was relatively bright at the time of the ACIS observation resulting in a (background subtracted) count rate of 4.42$\pm$0.01 counts/s (0.3-8.5 keV), which is still below the level that would cause pile-up. We extracted source events from an 8\arcsec segment of the 1-dimensional CC mode image centered at the pixel with the highest counts; we extracted background events from a 20\arcsec long segment with a 10\arcsec buffer on both sides of the source extraction. We then applied barycentric correction to all event arrival times.

We also observed \sgr with the {\it Chandra}/High Resolution Camera \citep[{\it HRC};][]{murr00} in Imaging mode for 10 ks on 2008 September 25 (MJD 54734). We used the superb angular resolution of the {\it HRC} to derive the precise X-ray position of the source. The X-ray intensity of the source was considerably lower at the time of the {\it HRC} observations with a (background subtracted) count rate of 0.505$\pm$0.007 counts/s in the 0.3$-$8.5 keV range.

\subsection{\it XMM-Newton} 

We observed \sgr with {\it XMM-Newton} on 2008 August 23 (MJD 54701) for 48.5 ks. Here, we used data collected with the European Photon Imaging Camera \citep[{\it EPIC};][]{pfef99} PN camera operated in Imaging with Prime Small Window mode. We selected source events from a circular region with a radius of 20\arcsec centered at the source and background events from a source-free region on the same chip. The background subtracted source count rate is 5.62$\pm$0.01 counts/s in the 0.3$-$8.5 keV range. Inspection of the source light curve revealed many short SGR events, as expected since the observation took place during the burst active episode of the source. We adopted our burst finding algorithm to 
search in the 0.3$-$10 keV band of the {\it EPIC-PN} data and identified 642 short bursts over a wide range of X-ray intensities. After removing the time intervals corresponding to all bursts, the exposure time was reduced to 47.5 ks. Finally we applied a barycentric correction to the event arrival times for our  timing studies.

\section{Data Analyses and Results} 

\subsection{Source Location}

We generated an image of the entire {\it Chandra}/{\it HRC-I} field in the 0.5$-$7.0 keV range and used wavdetect\footnotemark \footnotetext{a CIAO tool, http://cxc.harvard.edu/ciao/} to identify all X-ray sources in our frame. We found two point sources: \sgr at the aim point of the image and a previously uncatalogued X-ray source (CXOU\,J$050111.3+451525$) that is separated by 1.4$\arcmin$ from the SGR location. Their locations are 
RA = $05^h01^m06^s.76$, Dec = $+45\arcdeg16\arcmin33.92\arcsec$ (J2000), and RA = $05^h01^m11^s.33$, Dec = $+45\arcdeg15\arcmin25.28\arcsec$ (J2000) for \sgr and CXOU J$050111.3+451525$, respectively. 

The statistical uncertainty of the X-ray location of the latter source is 0.09\arcsec (22 source counts). To derive a precise positional uncertainty, we searched the 2MASS catalogue and found an IR source (2MASS $05011132+4515252$) at RA = $05^h01^m11^s.33$, Dec = $+45\arcdeg15\arcmin25.25\arcsec$ with an uncertainty of 0.06\arcsec. This location is consistent within the uncertainties, with the {\it Chandra} location of CXOU J$050111.3+451525$, so we concluded that the latter is very likely the X-ray counterpart to 2MASS $05011132+4515252$. Given the excellent alignment between these two localizations, we estimate the absolute positional uncertainty of \sgr to be 0.11$\arcsec$ (1$\sigma$).

\subsection{Pulse Properties}

\subsubsection{Pulse Timing Analysis}

\begin{figure}
\vspace{0.0in}
\centerline{
\epsscale{1.3}
\plotone{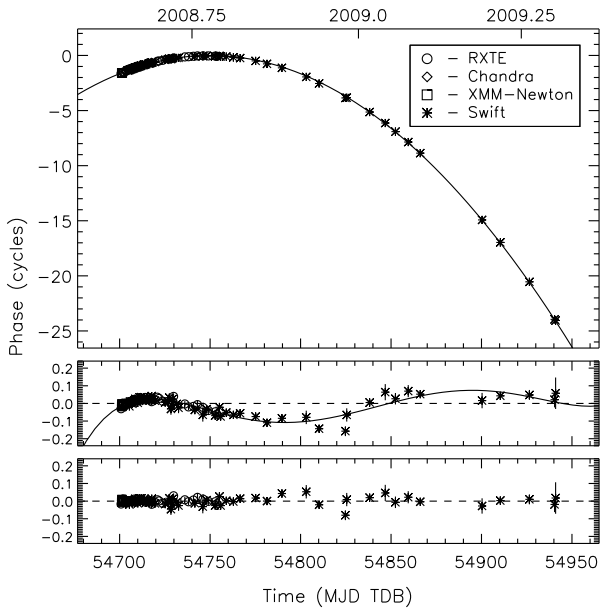}}
\vspace{-0.7in}
\caption{{\it Top panel} Plot of pulse phase drifts of each \sgr observation with respect to the pulse template (see also the text); {\it Middle Panel} the fit residuals when the pulse phase evolution is modeled with a fifth order polynomial; {\it Bottom Panel} the fit residuals when a quadratic spline fit is applied.
\label{fig:poly_spline_fit}}
\end{figure}

For this analysis we used the pulse timing technique described in detail in \citet{woods02}. In summary, we use an epoch folding algorithm to determine the pulse frequency and its higher order time derivatives as follows. We first group all available observations ({\it RXTE/PCA, Swift/XRT, Chandra/ACIS, XMM-Newton/EPIC-PN}) into discrete time segments. We then generate a (high signal to noise ratio) pulse profile template using the data in the $2-10$ keV energy band by accumulating multiple, closely spaced pointings. Next we cross-correlate the pulse profile obtained by each group of pointings with the template profile and measure the phase drifts with respect to the template (see Figure \ref{fig:poly_spline_fit}, top panel). We obtained a good fit to the phase residuals by modeling with a 5th order polynomial ($\chi$$^2$/degrees of freedom = 139.5/110). In Table \ref{tab:poly_fit} we provide the best fit parameters of the spin ephemeris of \sgrnos. The fit residuals are presented in the middle panel of Figure \ref{fig:poly_spline_fit}. We also employed a quadratic spline fit to all available phase residuals in seven segments. In Table \ref{tab:spline_fit} we list the spin frequencies and their time derivatives as determined with the quadratic spline method; the bottom panel of Figure \ref{fig:poly_spline_fit} shows the phase residuals obtained with the spline fit. It is important to note that in the pulse timing technique we employ, a systematic pulse profile change in which the profile drifts in phase cannot be distinguished from phase drift due to timing noise.

\begin{table}[h]
\caption{Pulse ephemeris of \sgr obtained using the combined {\it RXTE/PCA, Swift/XRT, CXO/ACIS-S}
and {\it XMM-Newton/EPIC-PN} observations}
\centering
\begin{tabular}{lc}\hline\hline
Parameter                   & Value   \\ \hline
Range (MJD)  &  54700.794 $-$ 54940.953 \\
Epoch (MJD)  	  & 54750.0  \\
$\nu$ (Hz)        & 0.173547943(1) \\
$\dot{\nu}$ ($10^{-13}$ Hz s$^{-1}$)  &  -1.752(8) \\
$\ddot{\nu}$ ($10^{-21}$ Hz s$^{-2}$) &   6.9(3) \\
$\nu^{(3)}$ ($10^{-27}$ Hz s$^{-3}$)  &  -2.7(3) \\
$\nu^{(4)}$ ($10^{-34}$ Hz s$^{-4}$)  &   3.5(6) \\ \hline
\end{tabular}
\label{tab:poly_fit}
\end{table}
 
\begin{table}[h]
\caption{Pulse frequency and frequency derivative of \sgr obtained with a 
quadratic spline fit to the {\it RXTE/PCA}, {\it Swift/XRT}, {\it Chandra/ACIS-S} and {\it XMM-Newton/PN} data}
\centering
\begin{tabular}{cccc}\hline\hline
Epoch   &  Range       &  $\nu$   &  $\dot{\nu}$  \\ 
(MJD)   &  (MJD)       &  (Hz)    &  ($10^{-13}$ Hz s$^{-1}$) \\ \hline
54707.5 & 54700.0 $-$ 54715.0 & 0.173548659(2) &  -2.03(8)  \\
54722.5 & 54715.0 $-$ 54730.0 & 0.173548384(3) &  -2.21(8)  \\
54740.0 & 54730.0 $-$ 54750.0 & 0.173548100(3) &  -1.62(8)  \\
54765.0 & 54750.0 $-$ 54780.0 & 0.173547721(4) &  -1.85(6)  \\
54805.0 & 54780.0 $-$ 54830.0 & 0.173547146(3) &  -1.55(4)  \\
54855.0 & 54830.0 $-$ 54880.0 & 0.173546394(4) &  -1.93(4)  \\
54915.0 & 54880.0 $-$ 54950.0 & 0.17354545(2)  &  -1.75(6)  \\  \hline
\end{tabular}
\label{tab:spline_fit}
\end{table}

\subsubsection{Pulse Profile Evolution}

To investigate the pulse morphology and its evolution in energy and time, we generated pulse profiles in the 0.3$-$2 keV, 2$-$4.5 keV, and 4.5$-$8.5 keV energy intervals for the {\it Chandra/ACIS-S}, {\it XMM-Newton/EPIC-PN} and {\it Swift/XRT} data, and in the 2$-$4.5 keV, 4.5$-$8.5 keV, 8.5$-$14 keV and 14$-$40 keV intervals for the {\it RXTE/PCA} data. We used the phase connected spin ephemeris presented in Table \ref{tab:poly_fit} for determining the phases of all event arrival times in each specified energy interval.

\begin{figure}
\vspace{0.0in}
\centerline{
\plotone{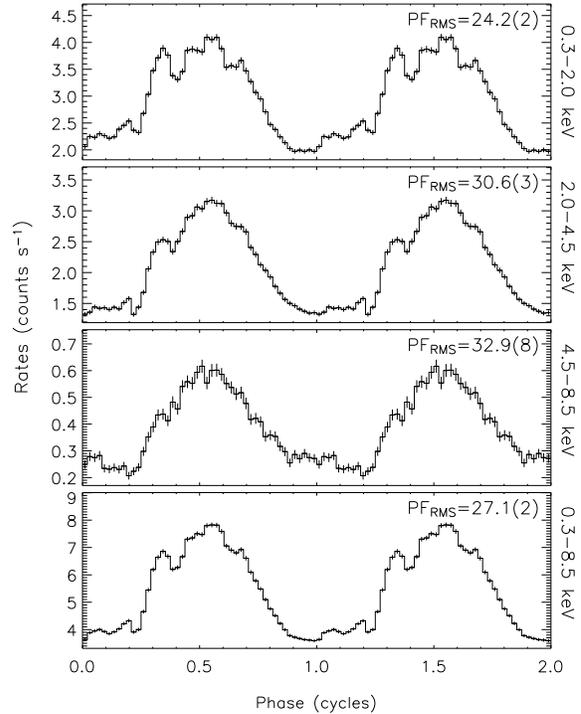}}
\vspace{0.1in}
\caption{{\it XMM-Newton/EPIC-PN} pulse profiles of \sgr with their  RMS pulsed fraction values  in various energy ranges as indicated on each panel. 
\label{fig:xmm_pp}}
\end{figure}

Figure \ref{fig:xmm_pp} shows the pulse profile evolution with energy as observed with {\it XMM-Newton/EPIC-PN} on the day after the onset of activation and during an episode of continued bursting. We find that the pulse profile in the 0.3$-$2 keV band is dominated by a broad peak ($\phi$$\sim$0.25$-$0.75) and a broad valley 
($\phi$$\sim$0.75$-$1.25), with overlying sub-pulses. The most significant sub-pulse appears at $\phi$$\sim$0.34. The significance of these sub-pulses decreases at higher energies and the profile becomes almost sinusoidal in the 4.5$-$8.5 keV range. The bottom panel in Figure \ref{fig:xmm_pp} demonstrates that the lowest energy interval emission is the most dominant component of the overall profile. 

The RMS pulsed fraction\footnotemark 
\footnotetext{The RMS pulsed fraction is defined as
${\rm PF}_{\rm RMS} = \left( \frac{1}{\rm N} ( \sum_{i=1}^{\rm N}
({\rm R}_{\rm i}-{\rm R}_{\rm ave})^2 -
\Delta {\rm R}_{\rm i}^2) \right)^\frac{1}{2}$ / ${\rm R}_{\rm ave}$
, where N is the number of phase bins (N=24), ${\rm R}_{\rm i}$ is the rate in each phase bin, $\Delta {\rm R}_{\rm i}$ is the associated uncertainty in the rate, and ${\rm R}_{\rm ave}$ is the average rate of the pulse profile.}
is energy dependent and increases with energy: (24.2$\pm$0.2)\%, (30.6$\pm$0.3)\%, (32.9$\pm$0.8)\% in 0.3$-$2 keV, 2$-$4.5 keV, 4.5$-$8.5 keV, respectively. The RMS pulsed fraction in the entire energy range (0.3$-$8.5 keV) is (27.1$\pm$0.2)\%.

\begin{figure}
\vspace{0.0in}
\centerline{
\plotone{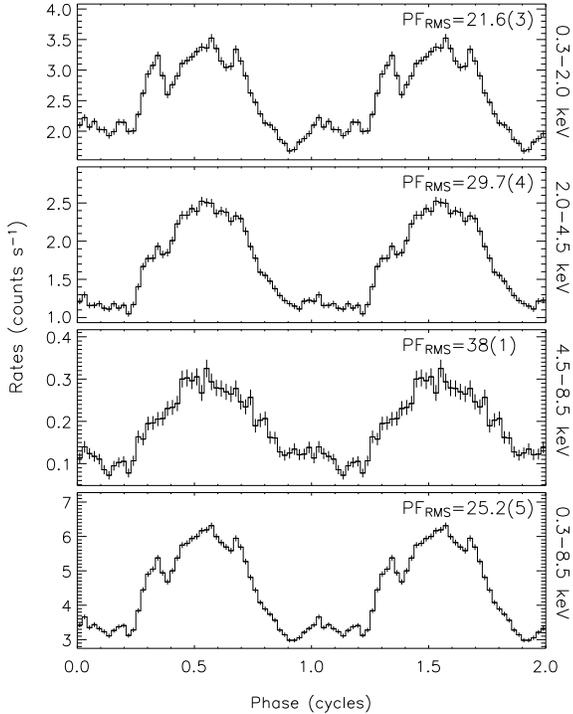}}
\vspace{0.1in}
\caption{{\it Chandra/ACIS-S} pulse profiles of \sgr with their  RMS pulsed fraction values in various energy ranges as indicated on each panel.
\label{fig:cxo_pp}}
\end{figure}

In Figure \ref{fig:cxo_pp}, we show the pulse profiles of \sgr seen with {\it Chandra/ACIS-S}  near the end of the burst active episode of the source. We find that the morphology of these profiles closely resemble in all energy ranges those obtained with the {\it XMM-Newton}, despite important differences in the source state: (i) the intensities (count rates) in all energy ranges are lower, (ii) the sub-pulse that peaks around $\phi$$\sim$0.03 becomes more prominent, and (iii) some pulsed fractions are significantly different. Most prominently, the {\it Chandra} 0.3$-$2 keV RMS pulse fraction is (21.6$\pm$0.3)\%, which is 7.2$\sigma$ lower than that seen with {\it XMM-Newton}. The 2$-$4.5 keV range pulse fraction is (29.7$\pm$0.4)\% (consistent with the {\it XMM-Newton} data within error). Finally the 4.5$-$8.5 keV band has an RMS pulsed fraction of (38$\pm$1)\%, which is 4$\sigma$ higher than that in the {\it XMM-Newton} data. The energy-averaged RMS pulsed fraction (0.3$-$8.5 keV) is (25.2$\pm$0.3)\%.

\begin{figure*}[h]
\vspace{0.0in}
\centerline{
\epsscale{1.1}
\plotone{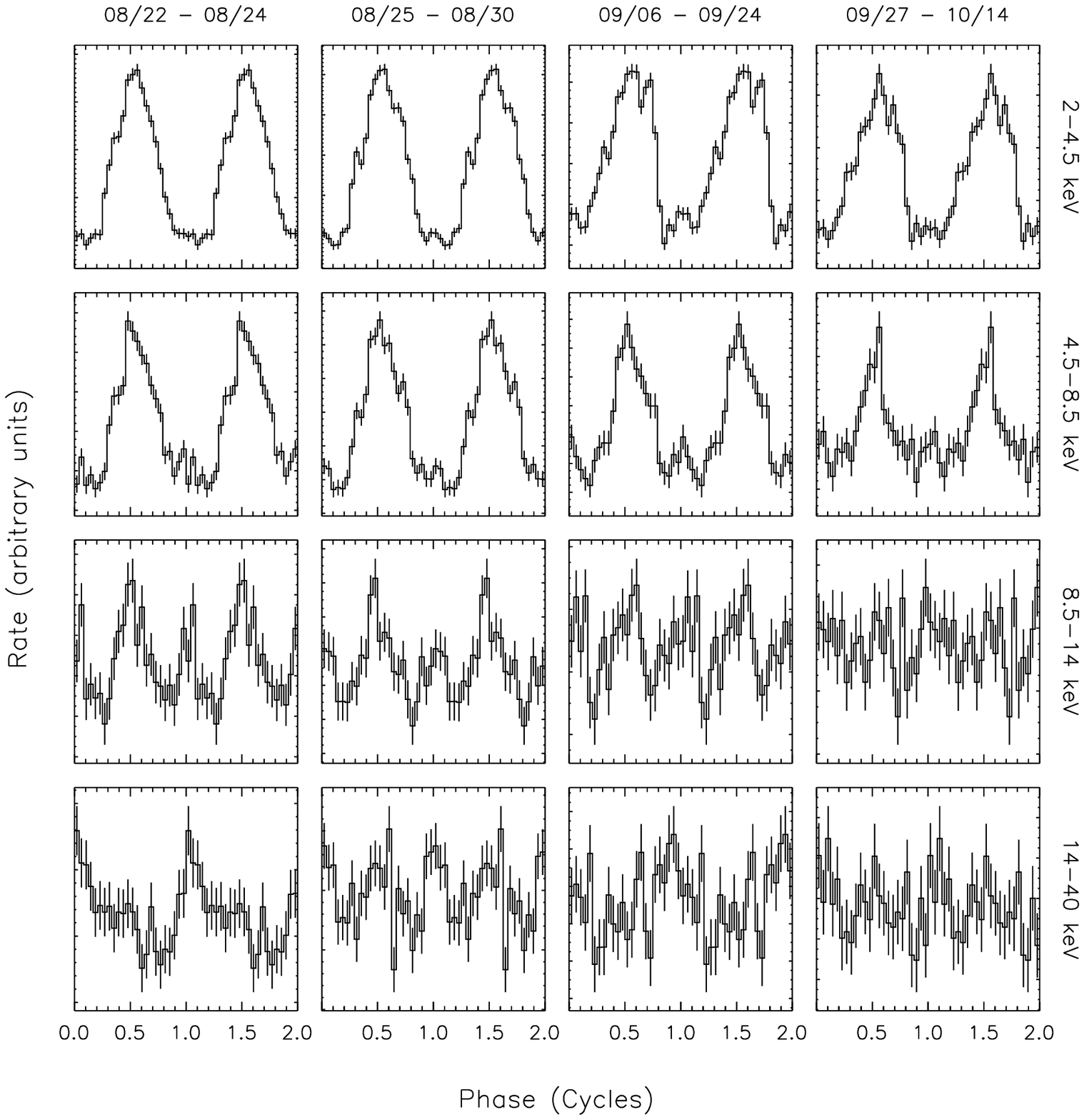}}
\vspace{0.1in}
\caption{Pulse profile history  of \sgr in the energy intervals shown on the right of each 
horizontal panel, obtained with {\it RXTE/PCA}. 
The labels on top of columns are the time ranges (in MM/DD format) within which the data 
used to construct these profiles were accumulated.
\label{fig:pca_pp}}
\end{figure*}

We present in Figure \ref{fig:pca_pp} the {\it RXTE/PCA} energy-resolved pulse profiles and their evolution in time. For this analysis, we have split the {\it PCA} observations into 4 groups of nearly equal exposures and obtained pulse profiles for each in the instrument energy range of $2-40$ keV. As the {\it PCA} is not an imaging instrument, its background determination relies on model based estimates and most likely does not reflect the true background; for that reason, we plot the pulse profiles in arbitrary units.

The leftmost column of Figure \ref{fig:pca_pp} corresponds to pulse profiles obtained during the most burst active episode of the source ($\sim$75\% of bursts were emitted in the two days following the episode onset). The pulse profile in the 2$-$4.5 keV range is characterized by a single, smooth pulse with a large duty 
cycle ($\sim$60\%). In the 4.5$-$8.5 keV band, the pulse structure is still broad and peaks around $\phi$$\sim$0.5, as in the lower energy interval, but it is asymmetric. Note that there is also evidence of weaker structures around $\phi$$\sim$1.0. The pulse profile in the 8.5$-$14 keV band is multi-peaked: the structure around $\phi$$\sim$1.0 becomes more prominent and almost as significant as the other pulse in the profile at about half the spin cycle. However, only the structure around $\phi$$\sim$1.0 remains in the 14$-$40 keV energy band.

In the time interval from 2009 August 25 to 30, (the second column from the left in Figure \ref{fig:pca_pp}), the source was at a very low level of bursting activity. Here we see again the structure at around $\phi$$\sim$1.0 and the sub-pulse near $\phi$$\sim$0.34 that was also seen in the 2$-$4.5 keV band profiles of {\it XMM-Newton} and {\it Chandra}. In the higher energy bands there are no remarkable changes in pulse shapes compared to the earlier interval, except for the fact that the structure near $\phi$$\sim$1.0 is now less significant. 

The last short burst from \sgr was observed on 2009 September 3, therefore, the profiles corresponding to the time intervals from September 6 to 24, (third column from the left in Figure \ref{fig:pca_pp}), and from September 27 to October 14, (the rightmost column in Figure \ref{fig:pca_pp}), are characteristic of the source's descent to the quiescent phase. In the 2$-$4.5 keV band of the third column we find a sub-pulse around $\phi$$\sim$0.7 that was significantly seen in the 0.3$-$2 keV {\it Chandra} profile. The structures peaking at $\phi$$\sim$0.5 and $\phi$$\sim$1.0 are clearly seen in the 4.5$-$8.5 keV band, while they are marginally seen in the 8.5$-$14 keV range. The pulse profile of the highest energy interval only shows marginal evidence of the structure near $\phi$$\sim$1.0. The 2$-$4.5 keV pulse profile of the rightmost column contains a single structure with a large duty cycle, broadly resembling profiles in the same band of earlier intervals. We find that the duty cycle of this pulse is significantly lower ($\sim$0.35 of the spin phase) in the 4.5$-$8.5 keV energy band. The pulse profiles of the higher energy bands of this episode are consistent with random fluctuations. 

\begin{figure*}
\vspace{0.0in}
\centerline{
\epsscale{1.1}
\plotone{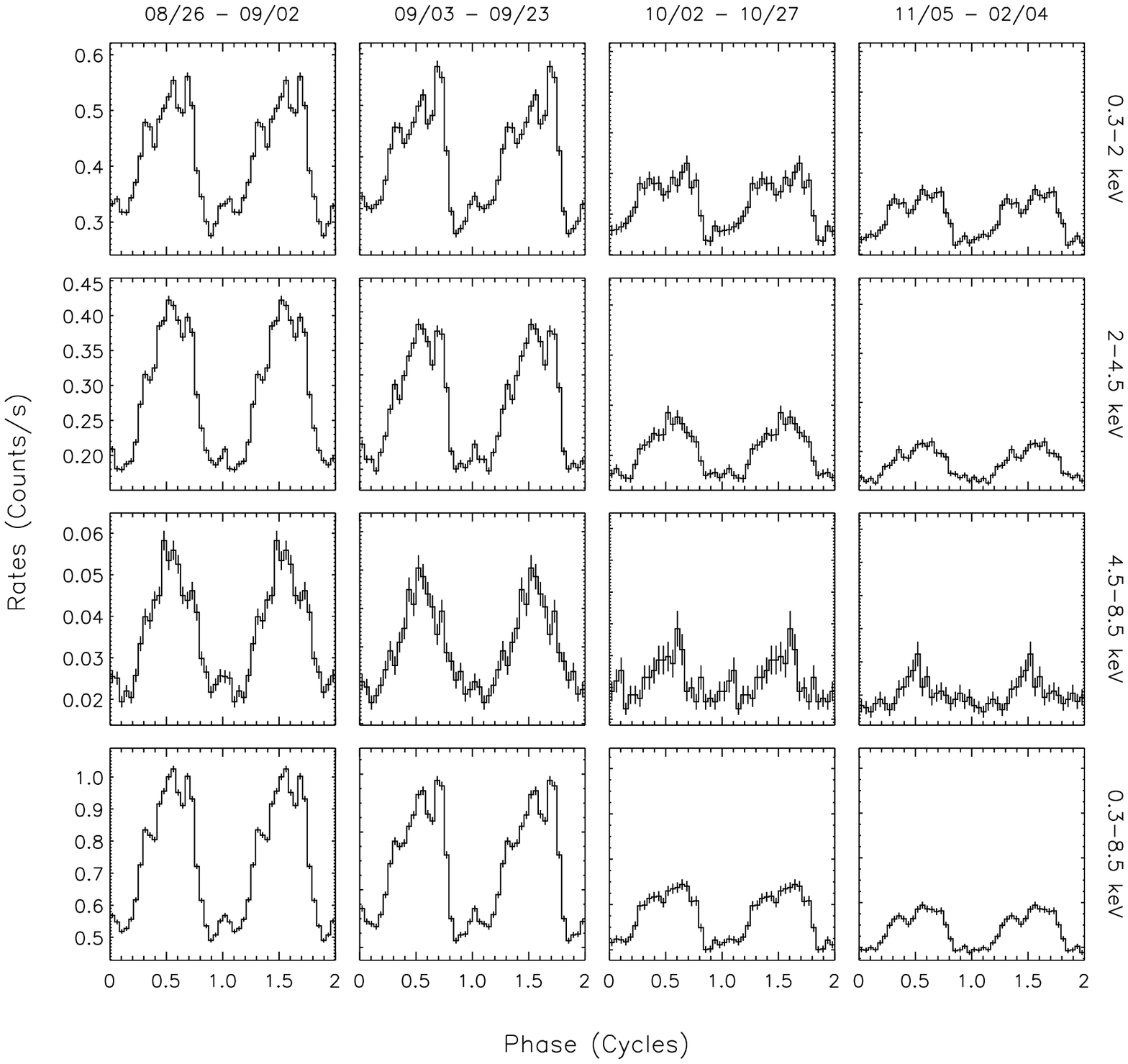}}
\vspace{0.1in}
\caption{Pulse profile history  of \sgr in the energy intervals shown on the right of each 
horizontal panel, obtained with {\it Swift/XRT}. The labels on top of columns are the time ranges (in MM/DD format) within which the data used to construct these profiles were accumulated. Note that the year changes to 2009 at the rightmost column.
\label{fig:xrt_pp}}
\end{figure*}

Finally, we also studied the \sgr energy and time resolved pulse profile evolution using the {\it Swift/XRT} observations which span a much longer time. We grouped the {\it XRT} pointings into four segments, each of which is comprised of nearly equal exposure time. The first interval corresponds to the burst active episode of the source, (the leftmost column in Figure \ref{fig:xrt_pp}), the second interval starts immediately after bursting ceased, (the second column from left in Figure \ref{fig:xrt_pp}), while the third and fourth intervals correspond to the quiescent state (long after the burst active phase; the observation end date is 2009 February 4). Unlike the {\it RXTE/PCA} pulse profiles in Figure \ref{fig:pca_pp}, we plot the {\it Swift/XRT} profiles in real units since proper background subtraction is possible with the XRT data. We find that the sub-pulses seen in the low energy bands of the {\it XMM-Newton} and {\it Chandra} observations persist during the burst active episode and in the interval immediately after, but disappear during quiescence. Interestingly, the sub-pulse near $\phi$$\sim$0.7 becomes remarkably prominent in the 0.3$-$2 keV and 2$-$4.5 keV ranges within days as the bursting stops. The pulses profiles of the interval of 40$-$65 days after the outburst onset reveal that the pulsed amplitudes in all energy ranges were significantly reduced (the third column from left in Figure \ref{fig:xrt_pp}) and sub-pulses seen in the lower energies have disappeared. During the quiescent phase (the rightmost column in Figure \ref{fig:xrt_pp}), we find further gradual decline in pulsed intensity, while similar to the previous time interval, we see only one significant pulse with duty cycle decreasing with energy.

\begin{figure}
\vspace{0.1in}
\centerline{
\epsscale{1.1}
\plotone{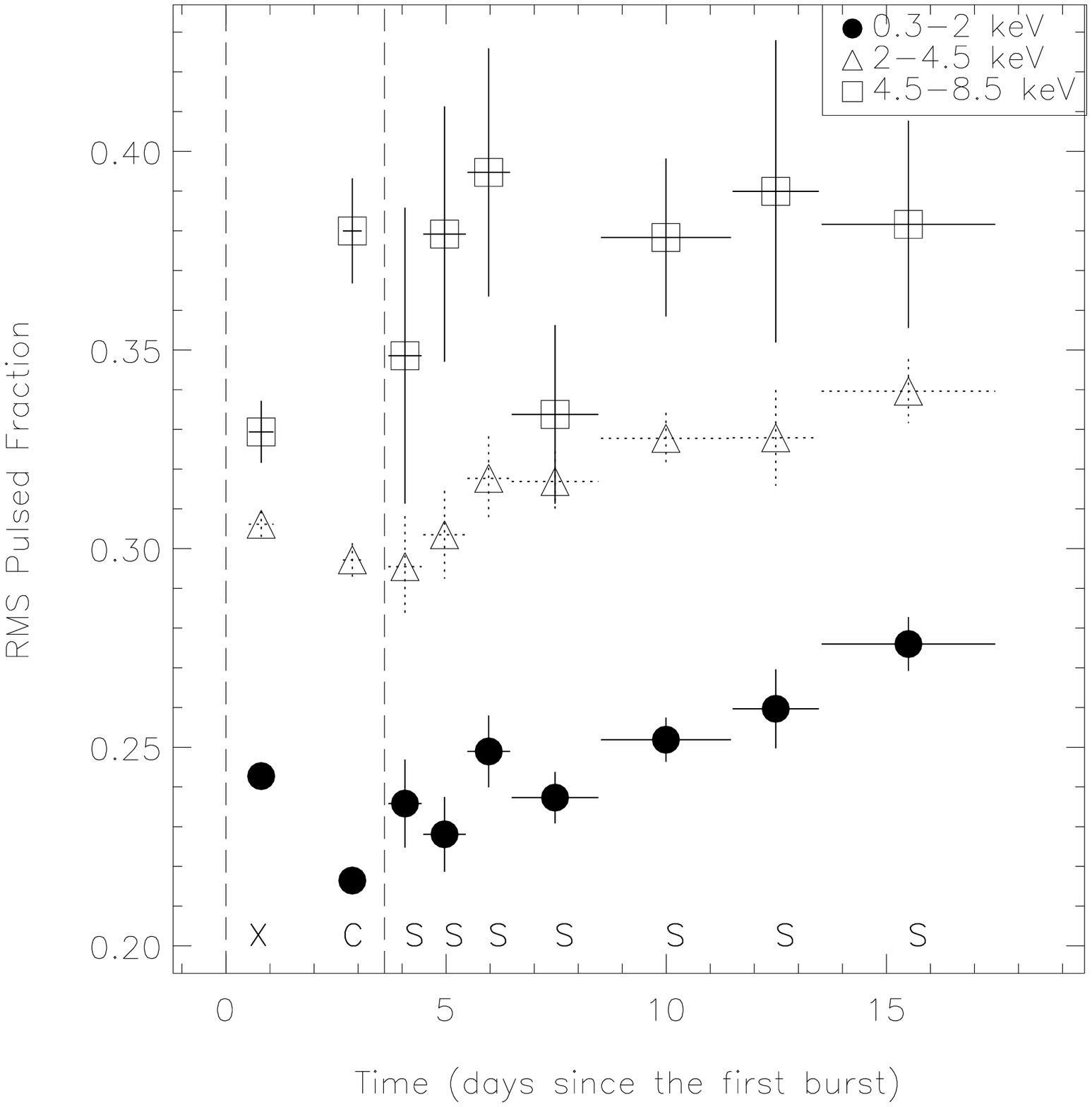}}
\vspace{0.3in}
\caption{Energy resolved RMS pulsed fraction variations of \sgr as obtained with the data of
{\it XMM-Newton/EPIC-PN, Chandra/ACIS-S} and {\it Swift/XRT}. The vertical 
dashed lines indicate the interval during which multiple short SGR bursts were detected. 
Letters at the bottom of the Figure indicate the instrument with which the plotted data above were obtained; X: {\it XMM-Newton}, C: {\it Chandra} and S: {\it Swift/XRT}.
\label{fig:rms_pf}}
\end{figure}

We plot in Figure \ref{fig:rms_pf}, the energy resolved variations of the RMS pulsed fractions during the burst active episode and immediately afterwards in the 0.3$-$2 keV, 2$-$4.5 keV and 4.5$-$8.5 keV intervals with {\it XMM-Newton, Chandra} and {\it Swift/XRT}. In the 0.3$-$4.5 keV band the pulsed fraction drops to its minimum value during ongoing source bursting. The lowest energy band RMS increases with time, while in the other two bands the deviation from a constant trend is less than 2.4$\sigma$.

\subsubsection{Pulsed Intensity Properties} 

\begin{figure}
\vspace{0.0in}
\centerline{
\plotone{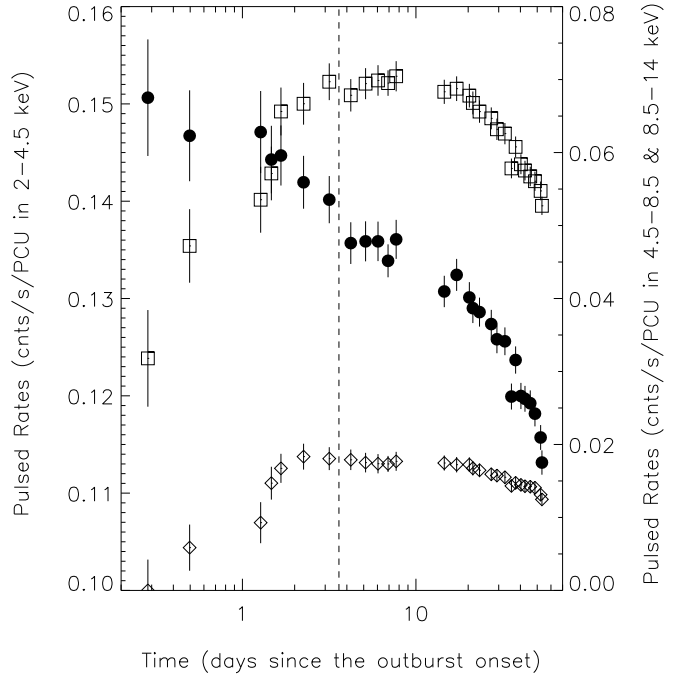}}
\vspace{0.1in}
\caption{Time history of pulsed count rates in the 2$-$4.5 keV (filled
circles), 4.5$-$8.5 keV (squares) and 8.5$-$14 keV (diamonds). 
Note that the tick mark values on the left axis correspond to the
2$-$4.5 keV band data, while the values on the right axis to the higher
two energy bands. The vertical dashed line indicates the end of the
active bursting episode of the source.
\label{fig:pca_pcr}}
\end{figure}

To investigate the pulsed intensity variations of \sgr in time and energy, we employed the RMS pulsed count rate\footnotemark
\footnotetext{
${\rm PCR}_{\rm rms} = \frac{1}{\rm N} \left(\sum_{i=1}^{\rm N}
({\rm R}_{\rm i}-{\rm R}_{\rm ave})^2 -
\Delta {\rm R}_{\rm i}^2\right)^\frac{1}{2}$
with propagated errors
$\delta{\rm PCR_{\rm rms}} = \frac{1}{{\rm N} ~ {\rm PCR}_{\rm rms}}
\left(\sum_{i=1}^{\rm N}
\left[({\rm R}_{\rm i}-{\rm R}_{\rm ave})
\Delta {\rm R}_{\rm i}\right]^2\right)^\frac{1}{2}$
}, 
which is not affected by background, therefore provides a measure of the pulsed intensity. We used the {\it PCA} observations since its exclusive coverage starts from soon after the onset of the source activation, continues throughout the active episode and extends well into the burst quiescent phase, while the {\it Swift/XRT}  observations start after the end of the major bursting activity. We present in Figure \ref{fig:pca_pcr} the evolution of the RMS pulsed intensity in
three energy intervals. We find that the pulsed intensity in the 2$-$4.5 keV band gradually declines while the source is actively emitting bursts, while the pulsed intensities in the 4.5$-$8.5 keV and 8.5$-$14 keV bands increase at the same time. The pulsed intensities remained constant in all energy bands for about a week following the active bursting phase, and then declined. The steepness of the decline is inversely proportional to the energy range.

\begin{figure}
\vspace{0.0in}
\centerline{
\plotone{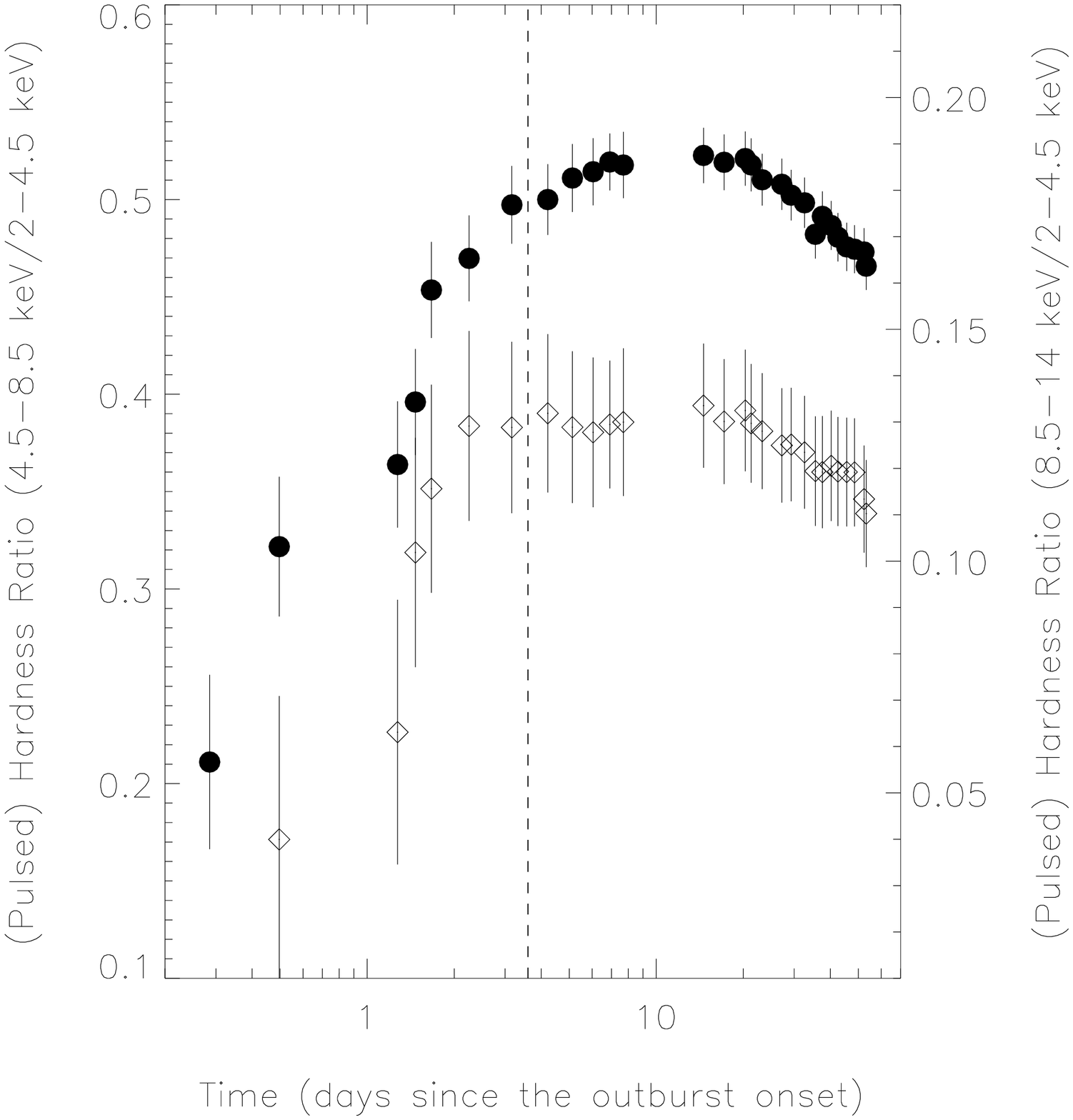}}
\vspace{0.1in}
\caption{Time history of hardness ratios of RMS pulsed intensities
in the 4.5$-$8.5 keV to that in the 2$-$4.5 keV (filled circles)
and in the 8.5$-$14 keV to that in the 2$-$4.5 keV 
(diamonds). The vertical dashed line indicates the end of
active bursting episode of the source.
\label{fig:pca_hr}}
\end{figure}

In Figure \ref{fig:pca_hr} we show the evolution of the pulsed hardness ratios, which are the ratios of the pulsed intensity in the two higher energy bands to that in the 2$-$4.5 keV band. We find that both ratios increase rapidly in the first 1.5 days into the outburst, continue to increase at a slower pace until the end of the burst active phase, remain constant for about a week, and decline afterwards.

\subsection{Spectral Properties}

\begin{figure}
\vspace{0.1in}
\centerline{
\plotone{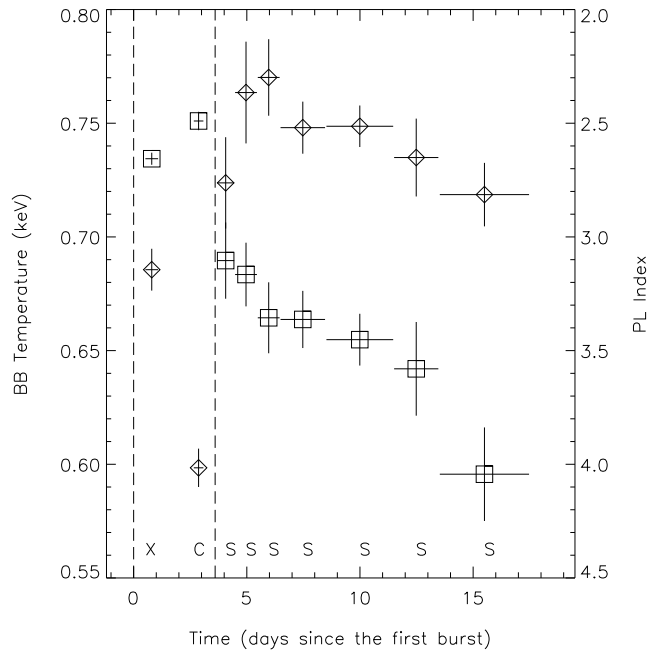}}
\vspace{0.3in}
\caption{Evolution of the X-ray spectral parameters of \sgrnos: blackbody temperature (diamonds) and power-law index (squares). The vertical dashed lines indicate the interval during which intense short SGR bursts were detected. Letters at the bottom of the Figure indicate the instrument with which the plotted data above were obtained; X: {\it XMM-Newton}, C: {\it Chandra} and S: {\it Swift}/XRT. 
\label{fig:spec_hist}}
\end{figure}

We analyzed the X-ray spectra of \sgr using the {\it XMM-Newton, Chandra/ACIS-S} and the {\it Swift/XRT} observations during and immediately after the burst active episode of the source. With respect to the longest extending observations of the {\it Swift/XRT}, we generated 7 combined spectra using multiple pointings to provide statistically significant spectral parameters and close in time, to avoid blending different spectral shapes (in case of spectral evolution). We fit all spectra simultaneously with an absorbed blackbody plus power-law model by linking the interstellar absorption parameter in the 0.5$-$10 keV energy range. We obtained an acceptable fit to all spectra ($\chi$$^2$/degrees of freedom = 5086.2/4470 = 1.14) with a hydrogen column density, $N_{\rm H}=$(0.95$\pm$0.08)$\times$10$^{22}$ cm$^{-2}$. In Figure \ref{fig:spec_hist} we present the evolution in time of our best fit blackbody temperature and power-law index. We find that the blackbody temperature declined very rapidly while the source was burst active, increased to a maximum value of $kT=0.77$ keV soon after the intense bursting ceased, and then gradually decreased. The power law component, on the other hand, became less steep during the burst active phase, and then steeper as the source went into burst quiescence.

\begin{figure}
\vspace{0.0in}
\centerline{
\epsscale{1.2}
\plottwo{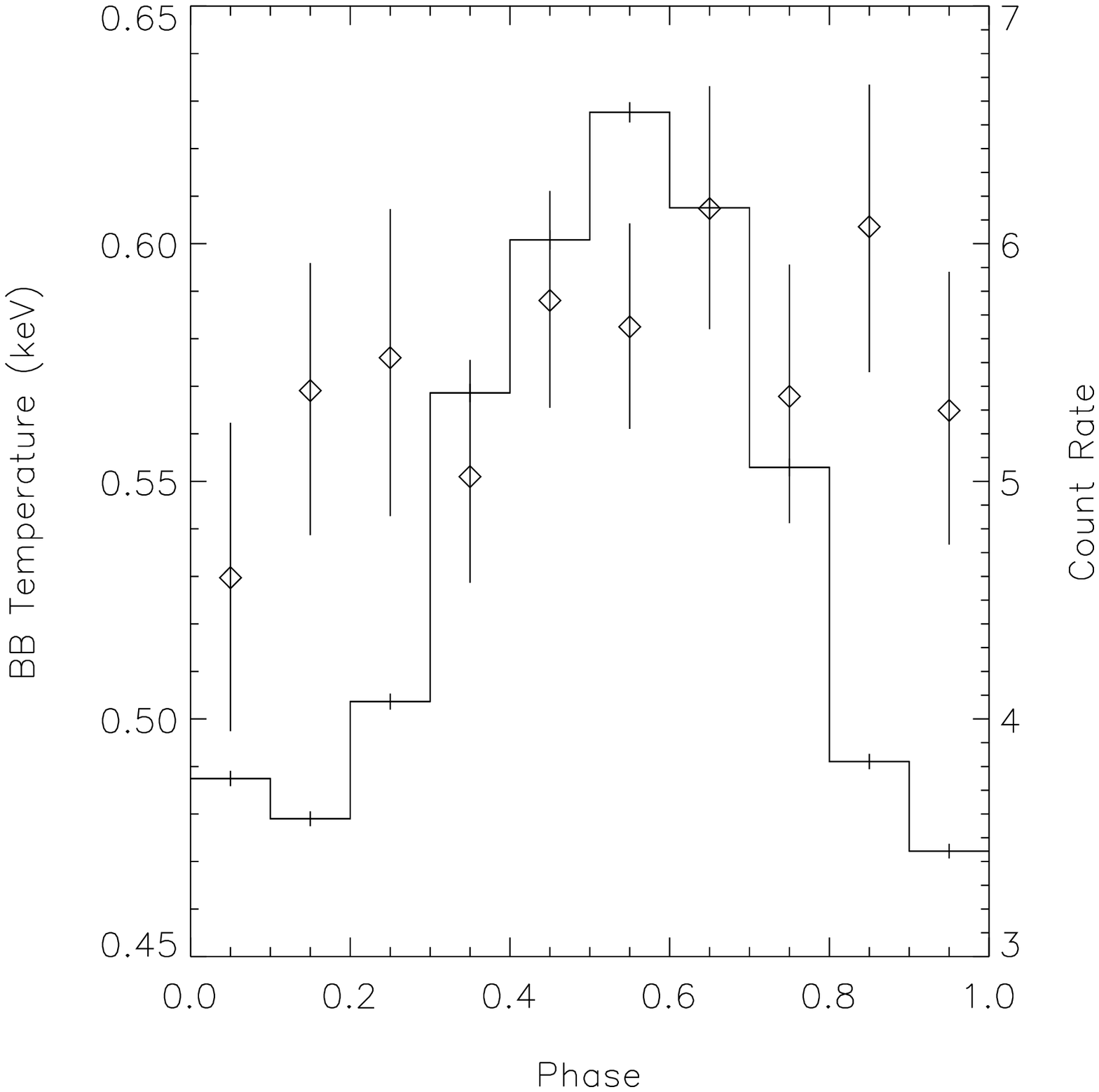}{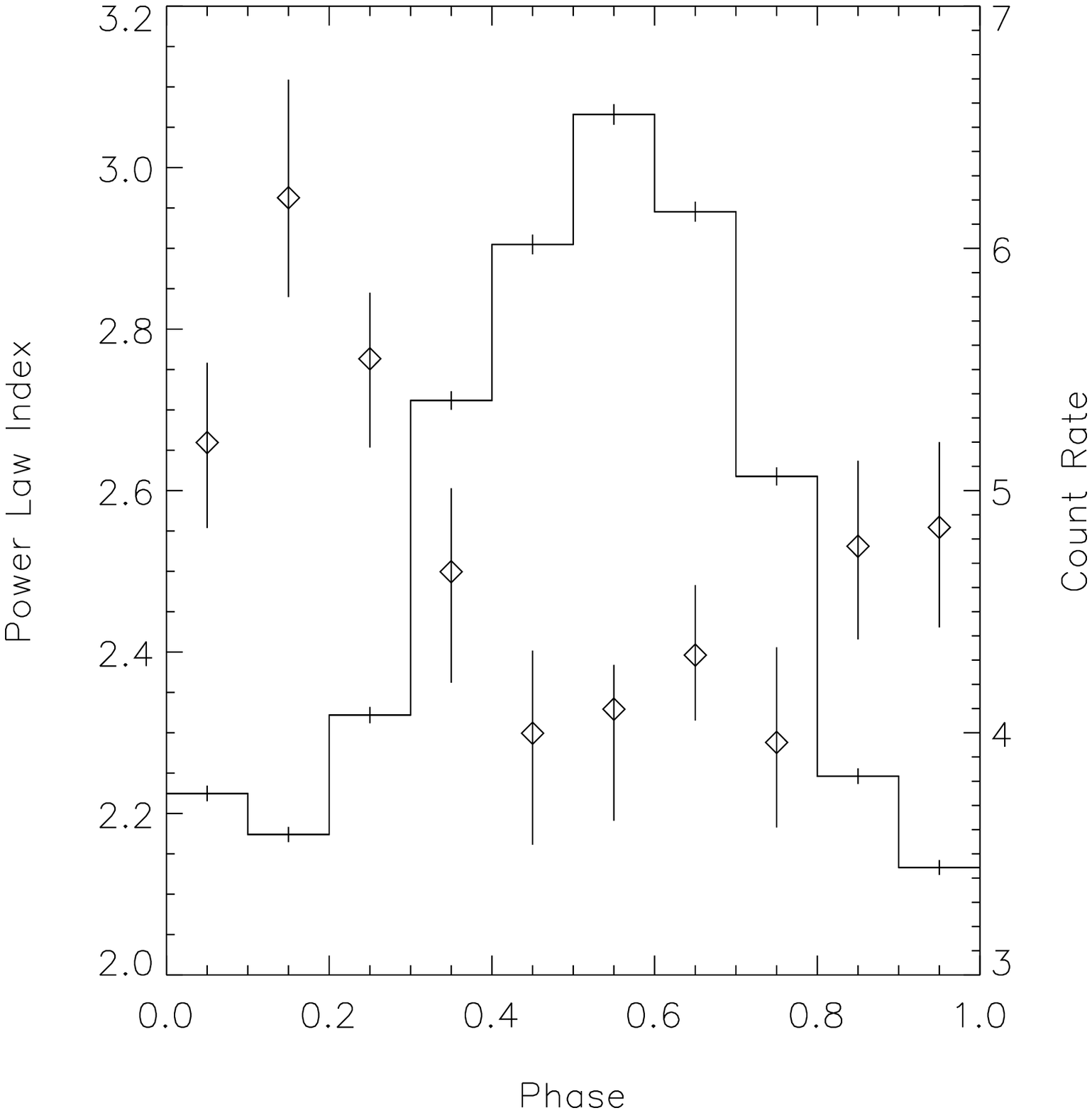}}
\vspace{0.3in}
\caption{Phase-resolved spectral parameters of \sgr measured using
{\it Chandra/ACIS-S} observations. {\it Left Panel:} blackbody temperatures; {\it Right Panel:}
power law indices. We overplot the pulse profile (histogram with units on the right) binned into phase steps of the phase resolved spectral extraction.
\label{fig:cxo_phres}}
\end{figure}

Motivated by the rapid spectral variations (as well as by the change in RMS pulsed fraction) seen in the {\it Chandra/ACIS-S} observations, we performed a phase-resolved spectral analysis of the data. As a reminder, the {\it Chandra} data cover the fourth day after the onset of the source burst activity. We divided the spin phase into 10 bins and extracted spectra for each bin. We then simultaneously fit all 10 spectra with the attenuated blackbody plus power-law model, again by linking the interstellar absorption. We obtained a good fit to all spectra ($\chi$$^2$/degrees of freedom = 1909.7/1785 = 1.06) with $N_{\rm H}=$(0.98$\pm$0.03)$\times$10$^{22}$ cm$^{-2}$, consistent with that obtained for the integrated spectra. In Figure \ref{fig:cxo_phres}, we present the best fit phase-resolved blackbody temperatures (left) and power law indices (right). We find that the blackbody temperature remains constant within errors over the whole spin phase, while 
the power law component is strongly intensity-dependent: the spectrum becomes softer during the phase minimum and harder during the phase maximum. 

Finally, to evaluate the variations in spectral shape and source flux as the source entered the quiescent phase, we generated 3 {\it Swift/XRT} spectra obtained with about 12$-$13 ks of exposures that took place long after the outburst episode (2009 October, December and 2010 February). We fit all three spectra simultaneously with the absorbed blackbody plus power-law model and with the hydrogen column density linked. We find that all three power-law indices are consistent with one another, therefore, we linked them as well to better constrain the blackbody component. We obtained a very good fit ($\chi$$^2$/degrees of freedom = 123.8/114 = 1.09) with a $N_{\rm H}=$(1.19$\pm$0.22)$\times$10$^{22}$ cm$^{-2}$, a common power-low index of 5.04$\pm$0.21 and blackbody temperatures of 0.78$\pm$0.07 keV, 0.59$\pm$0.07 keV and 0.56$\pm$0.06 keV, for the above mentioned three epochs, respectively. The corresponding 1$-$10 keV unabsorbed flux values were (7.0$\pm$0.3)$\times$10$^{-12}$, (6.4$\pm$0.2)$\times$10$^{-12}$ and (5.8$\pm$0.2)$\times$10$^{-12}$ erg cm$^{-2}$ s$^{-1}$.

\section{Discussion} 

\subsection{Persistent emission properties} 

We determined the pulse ephemeris of \sgr using the longest time baseline available ($\sim$240 days). We find that the spin ephemeris of the source can be described well with a fifth order polynomial (see Table 1) or by employing a quadratic spline method including only the spin frequency and frequency derivative (Table 2). We, therefore, argue that the recent report of $\ddot{P}$ being the recovery of the secular spin-down \citep{rea09} is not the only viable possibility. The two best studied SGR sources (SGR\,1806$-$20 and SGR\,$1900+14$) are known to exhibit torque noise \citep{woods02}; similarly, AXP\,1E 1048.1$-$5937 \citep{gk04}. Recently, \citet{belo09} suggested that untwisting of the magnetic field through the magnetosphere due to sudden crustal motion may account for the variable or noisy spin-down behavior of these sources. The variations seen in \sgr may then well be due to a noisy spin-down of the source.

We estimate the inferred dipole magnetic field strength of \sgr as $B_{\rm d}$ $\approx$ 2$\times$10$^{14}$ G, assuming that the neutron star slows down via magnetic braking. This value is very similar to the inferred field strength of the recently discovered SGR\,J1833$-$0832 \citep{gogus10} and lies near the lower end of the inferred magnetic fields of SGRs. The characteristic age (P/2$\dot{P}$) of \sgr is estimated to be 15.7 kyr. 
   
We find the energy and time dependence of the pulse profiles of \sgrnos, during its burst active episode and soon afterwards, intriguing. During the burst active phase, the $\sim$4.5 keV pulse profiles are smooth while those in the lower energy bands show significant subpulses. In particular, there are at least four subpulses between 0.3$-$2 keV, indicating that there are significant intensity variations over the spin phase. The most significant subpulse peaks near $\phi$$\sim$0.34 and lasts about 0.72 s, which corresponds to $\sim$1/8 of the spin period. 

The energy dependence of the pulse profile extends to the higher energies; the pulsations are clearly seen up to 40 keV following the onset of the outburst. The significant peak in the 14$-$40 keV band near $\phi$$\sim$1.0 is weaker in the 8.5$-$14 keV range and not seen in the lower energy bands. This hard component fades away within a few weeks after the burst activity onset. \citet{rea09} and \citet{enoto10} deduces the same decay behavior using {\it INTEGRAL} and {Suzaku/HXD} observations, respectively. A similar trend is also seen in the post-outburst pulse profiles of SGR\,J0418+5716 (P. M. Woods et al. 2010, in preparation). This fact may (indirectly) indicate the existence of heating of the neutron star crust as a result of crustal fracturing \citep{td96}.

We find that the pulsed fraction significantly increases with energy after a large drop in the 2$-$4.5 keV band near the end of the bursting episode. This trend is noteworthy but should be taken with caution since inter-instrument calibration may play a role as we discuss below.

\sgr went through significant spectral variations both during the bursting phase and throughout the quiescent phase, when no more SGR events were seen. We find that the spectrum of the source was hardening during the active period (see also \citet{rea09}). According to the magnetar scenario, this is a natural outcome of surface fracturing at a wide range of scales: sufficiently large scale crustal fracturing would lead to highly energetic short bursts \citep{td96} while smaller scale fracturing may contribute to the persistent X-ray emission of the source \citep{td95}. During the post-outburst episode, the spectrum of \sgr remained constant  for about a week and softened thereafter. This is also expected in the magnetar model as the heated stellar crust cools down \citep{guver07}.

The significant phase-resolved spectral variations near the end of the burst active phase seen with {\it Chandra} are very interesting. We showed that the blackbody temperature remains constant over the spin phase, while the power law component gets shallower (harder) with increasing X-ray intensity. Since the peak interval of the pulse phase would dominate the phase-averaged spectrum, it is possible to `hide' a shallow power law trend. On the other hand, the sharp drop in the blackbody temperature during the burst active phase and then its rapid rise in a day is indeed puzzling. Since this behavior was only seen with the {\it Chandra} data, there is no way to confirm these results with other contemporaneous observations.

Finally, we find that the flux in the 1$-$10 keV range measured with {\it Swift/XRT} between 412 and 546 days after the onset of the 2008 outburst,  follows a marginal decline from 7$\times$10$^{-12}$ to about 6$\times$10$^{-12}$ erg cm$^{-2}$ s$^{-1}$. It is interesting to note that the source flux in the same energy range at about 100 days after the onset was around 6$-$7$\times$10$^{-12}$ erg cm$^{-2}$ s$^{-1}$ (see Figure 11 of \citealt{rea09}). Therefore, the source flux likely remained constant over a year following a steady decrease over the first 100 days after activation.

\subsection{Archival outburst episodes of \sgr}

\begin{figure}
\vspace{-0.2in}
\centerline{
\epsscale{1.3}
\plotone{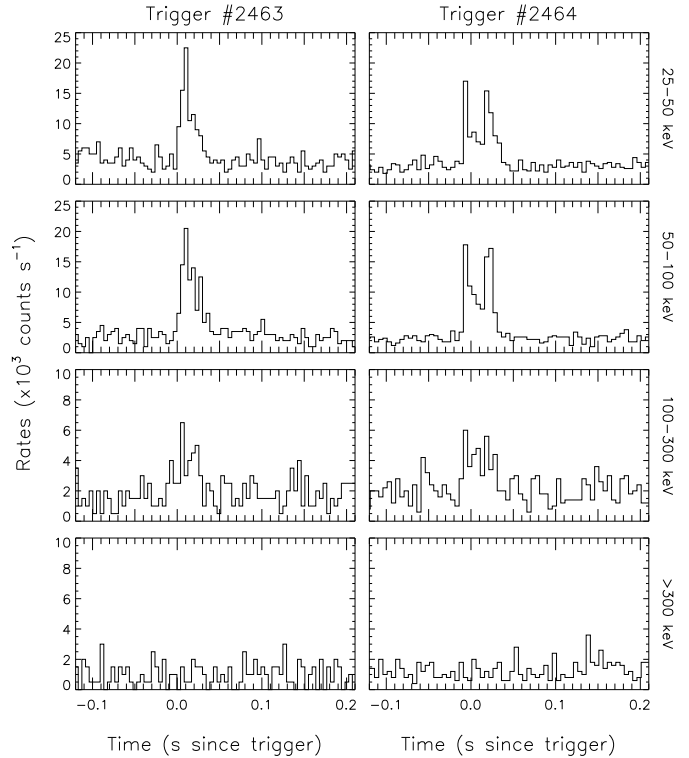}}
\vspace{0.1in}
\caption{Light curves of BATSE Trigger number 2463 (left) and 2464 (right) in the four LAD discriminator energy intervals which are indicated on the right. 
\label{fig:batse_lc}}
\end{figure}

On 1993 July 25, the Burst And Transient Source Experiment (BATSE) onboard the {\it Compton Gamma Ray Observatory (CGRO)} triggered on two peculiar events: at 17:45:21 UT (BATSE trigger number 2463) and at 23:32:37 UT (BATSE trigger number 2464). In Figure \ref{fig:batse_lc}, we present BATSE Large Area Detector (LAD) light curves of these two events in the four LAD discriminator energy intervals. These two bursts were both short (T$_{90}$ = 0.064 s and 0.048 s, respectively), had soft spectra and their locations were consistent with one another (RA, Dec= 05$^{\rm h}$ 09$^{\rm m}$ 12$^{\rm s}$, 43\degree 48$\arcmin$ 00$\arcsec$, error radius of 5.7\degree, and $04^{\rm h}$ 42$^{\rm m}$ 00$^{\rm s}$, 44\degree 12$\arcmin$ 00$\arcsec$, error radius of 6.2\degree, respectively). Note that the {\it Chandra} position of \sgr falls well within the overlap of these two error circles. Given the soft nature of these events, their short durations and the fact that they originated from the same region of the sky within about 5.8 hours apart, it is strongly suggestive that these events are from \sgr. Our search for untriggered events using the continuous BATSE DISCLA data (see \citealt{gogus99} for the methodology) for 10 days starting on 1993 July 20, resulted in no additional events from the source. Therefore, we conclude that \sgr was indeed burst active back in July 1993 for a very brief period.

Furthermore, the Gamma Ray Burst Monitor (GRBM) onboard the {\it BeppoSAX} satellite recorded a short burst (T$_{90}$ = 0.77 s) on 2000 October 11 at 10:06:49 UT (GRB 001011A) from RA = 04$^{\rm h} 28^{\rm m}$, Dec = 53\degree (with a positional uncertainty of 34\degree) that includes the position of \sgr \citep{front09}. Even though the spectrum of the GRBM burst is not very soft, there is another short duration event from the same direction about 95 s after the GRBM trigger. It is, therefore, suggestive that the source of these bursts is \sgr since they were positionally coincident, they were short and they repeated. Note, however, that the large error circle of GRBM also included another recently discovered magnetar source, SGR\,J0418+5729 \citep{vdhorst10}.

With the definite active episodes in 1993 and 2008, and a third possible one in 2000 we can infer that \sgr is going through burst active phases on a timescale of $\sim$10 years. This timescale is similar to that of SGR\,1627$-$41 which was discovered in 1998 during about a week-long burst active episode \citep{kouv98, woods99} and exhibited a burst active epoch again in 2008 \citep{espo08}. The two recently discovered SGRs (SGR\,J0418+5729 and SGR\,J1833$-$0832) have emitted only a few bursts during their very brief active episodes \citep{vdhorst10, gogus10} and their outburst recurrence behaviors may characteristically be different from \sgr and SGR\,1627$-$41.

\subsection{On the association with Supernova Remnant, G160.9+2.6}

\cite{gc2008} compared the position of SGR~0501+4516 with the locations \citep{gre09} of known Galactic supernova remnants (SNRs), and reported that the magnetar sits just outside the south-eastern rim of the well-known SNR~G160.9+2.6 \citep[also known as HB9][]{hh53,lt07}.

All of the four initially identified SGRs (1806--20, 0526--66, 1900+14, 1627--41) were originally claimed to sit near but outside SNRs \citep{cdt+82,abh+87,hlk+99,hsk+00}. If the SNR and neutron star in these cases are physically associated and hence formed in the same supernova explosion, the large angular offsets of the SGR positions from the centres of the respective SNRs require high space velocities of thousands of km~s$^{-1}$.  Indeed, it was hypothesized that magnetars received large kicks through magnetic field-driven mechanisms that were ineffective for ordinary pulsars \citep{dt92}.  However, in three of the four cases the density of SNRs on the sky in that region is sufficiently high that a chance projection between SGR and SNR cannot be ruled out, while for SGR 1806$-$20, the classification of the extended nebula as a SNR is now thought to have been erroneous \citep{gsgv01}. Recently, \citet{espo09} reported the detection of diffuse soft X-ray emission around SGR 1627$-$41 and interpreted it as emission from the supernova remnant, G337.0−0.1. Meanwhile, direct measurements of magnetar proper motions imply much lower velocities than inferred by these offsets \citep{hcb+07,kch+09}, compatible with the velocities observed for radio pulsars \citep{acc02,hllk05,cbv+09}.

Assuming that the progenitor supernova explosion occurred at the geometric center of the SNR, the SGR has traveled approximately 80~arcminutes from its birthplace over its lifetime. For a distance to the SNR of $800\pm400$~pc and an estimated age of 4000--7000 years \citep{lt07}, the implied projected space velocity for the SGR is 1300--6800 km~s$^{-1}$. As for the previous claimed SGR/SNR associations mentioned above, this is at or beyond the highest directly observed velocity observed for any other neutron star \citep[cf.][]{cvb+05,wp07}.  However, the proximity of this SNR allows a direct test: independent of distance or projection angle, if the SNR and SGR are physically associated, then the predicted proper motion of the SGR will be approximately 0.7--1.2 arcsec per year to the south, which should be measurable with the {\em Chandra X-ray Observatory}\ over even a relatively short time baseline. Confirming more associations would force a reconsideration of currently disfavored SGR--SNR associations, and of the highest velocities that neutron stars can attain.

\acknowledgments 

E.G. acknowledges the support from the Scientific and Technological Research Council of Turkey  (T\"UB\.ITAK) through grant 105T443. E.G. and Y.K. acknowledge EU FP6 Transfer of Knowledge Project “Astrophysics of Neutron Stars” (MTKD-CT-2006-042722). B.M.G. acknowledges the support of a Federation Fellowship from the Australian Research Council through grant FF0561298. {\it Chandra} observations were carried out under Observation ID 10164 and 9131, part of the proposal “ToO Observations of SGRs” (NASA grant GO9-0065Z; PI: C. Kouveliotou).

\end{document}